\documentclass{aastex631}
\usepackage{amsmath,amssymb}
\usepackage{natbib}

\shorttitle{Gamma-Ray Source toward 1H 1934-063}
\shortauthors{Li ET AL.}

\begin{document}

\title{Fermi-LAT Detection of a Gamma-ray Excess toward the Radio-quiet Narrow-line Seyfert 1 Galaxy 1H 1934--063}

\email{liyangji@ynao.ac.cn, baijinming@ynao.ac.cn}

\author{Yangji Li}
\affiliation{Yunnan Observatories, Chinese Academy of Sciences, Kunming 650216, China}
\affiliation{School of Astronomy and Space Science, University of Chinese Academy of Sciences, Beijing 100049, China}
\affiliation{University of Chinese Academy of Sciences, Beijing 100049, PR China}

\author{Jinming Bai}
\affiliation{Yunnan Observatories, Chinese Academy of Sciences, Kunming 650216, China}
\affiliation{University of Chinese Academy of Sciences, Beijing 100049, PR China}

\author{Xiong Jiang}
\affiliation{Key Laboratory of Dark Matter and Space Astronomy, Purple Mountain Observatory,
Chinese Academy of Sciences, Nanjing 210023, People’s Republic of China}
\affiliation{School of Astronomy and Space Science, University of Science and Technology of China, Hefei, Anhui 230026, People’s Republic of China}

\author{Kaixing Lu}
\affiliation{Yunnan Observatories, Chinese Academy of Sciences, Kunming 650216, China}
\affiliation{University of Chinese Academy of Sciences, Beijing 100049, PR China}

\keywords{galaxies: active --- galaxies: individual (1H 1934--063) --- galaxies: Seyfert --- gamma rays: galaxies}
\begin{abstract}

We report a $\gamma$-ray excess toward the radio-quiet narrow-line Seyfert~1 galaxy 1H~1934$-$063 using data collected by the Large Area Telescope (LAT) on board the \textit{Fermi} Gamma-ray Space Telescope and taking into account the LAT 16-year Source List (FL16Y). During the flare interval, the excess is detected in the 1--500~GeV band at a significance of $\sim5.2\sigma$ (TS = 27.12), with a photon flux of $(4.94\pm2.33)\times10^{-10}\ {\rm ph\ cm^{-2}\ s^{-1}}$ and a hard photon index of $\Gamma=1.50\pm0.25$. The best-fit $\gamma$-ray position is consistent with the radio position of 1H~1934$-$063, while the nearby source FL16Y~J1936.9$-$0552 is not significantly detected during the same interval. In the absence of contemporaneous multiwavelength data, the broadband interpretation cannot be tightly constrained. A compact nonthermal component can phenomenologically account for part of the GeV emission, but the present data do not allow a unique physical interpretation. The excess therefore provides an interesting case for probing high-energy activity in radio-quiet NLS1 galaxies, although the underlying physical mechanism remains uncertain.

\end{abstract}

\section{Introduction}
Narrow-line Seyfert 1 (NLS1) galaxies are a peculiar subclass of active galactic nuclei (AGNs), characterized by narrow permitted optical emission lines, relatively small black hole masses, and high accretion rates \citep{Osterbrock1985,Komossa2008}. A small but growing number of NLS1 galaxies have been established as $\gamma$-ray emitters \citep{Abdo2009PMNJ0948,Abdo2009RLNLS1,DAmmando2012SBS0846,DAmmando2015FBQSJ1644,Yang2018J2118,Li2023J0959,Foschini2022NewSample,Paliya2024NLS1Catalog}. These sources are predominantly radio-loud and are generally interpreted within a jetted framework \citep{Paliya2019,DAmmando2019}, showing that at least some NLS1 galaxies can produce strong nonthermal high-energy emission despite their modest black hole masses \citep{Abdo2009PMNJ0948,Abdo2009RLNLS1}. However, this picture is based mainly on radio-loud systems with clear jet signatures \citep{DAmmando2019}, and whether similar high-energy activity can arise in radio-quiet NLS1 galaxies remains uncertain.

The interpretation of $\gamma$-ray emission in radio-quiet NLS1 galaxies is correspondingly less straightforward. In jetted systems, the GeV output is usually attributed to inverse-Compton processes in relativistic outflows \citep{Blandford1979,Madejski2016,Boettcher2013,Boettcher2019}, whereas in radio-quiet AGNs other channels, including star formation, coronal activity, weak jets, and AGN-driven outflows, may also contribute \citep{Ackermann2012SFG,Inoue2023NGC4151,Sakai2025GRS1734}. Recent \textit{Fermi}-LAT studies have begun to populate this regime, both through stacked evidence for $\gamma$-ray emission from hot coronae in nearby radio-quiet Seyferts and through targeted analyses of individual Seyfert galaxies such as NGC 3281, for which a potential softer $\gamma$-ray component has been reported and may arise from either the corona or a jet-related component\citep{Liu2025NGC3281,Liu2025HotCoronae}. Recent theoretical work has likewise suggested that AGN accretion flows or disks may host compact embedded high-energy engines, including accretion-modified stars, compact-object accretion, and localized relativistic jets \citep{Wang2021AMSBBH,Liu2024AMSObs,Chen2023COAGN,Chen2025JetAGN}. These developments broaden the physical context in which high-energy activity in otherwise non-blazar AGNs may be interpreted.

In this context, 1H 1934$-$063 is a particularly interesting target. Located at R.A. = 294.3876$^\circ$ and decl. = $-6.2180^\circ$ (J2000), it is a nearby radio-quiet NLS1 at $z = 0.010382$, with a reported radio loudness of $R_{1.4}\simeq2.8$. Its radio properties are modest, with $S_{1.4\,\mathrm{GHz}}\approx 42.2$ mJy and $S_{5.5\,\mathrm{GHz}}\approx 12.1$ mJy; the 5.5 GHz image further shows diffuse kpc-scale emission around the central core and a steep radio spectrum \citep{Chen2020Radio}. The source is also included in the flux-calibrated 6dFGS southern NLS1 sample \citep{Chen2018South}. X-ray observations show that it is bright and rapidly variable, with evidence for reverberation and ionized outflows \citep{Frederick2018,Xu2022}. Against this background, a $\gamma$-ray excess toward 1H 1934$-$063 is noteworthy, although its association and physical origin require careful tests in the absence of contemporaneous multiwavelength data.

The remainder of this paper is organized as follows. Section~2 presents the \textit{Fermi}-LAT analysis and optical spectrum analysis. In Section~3, we compile a non-simultaneous broadband spectral energy distribution and summarize the relevant multiwavelength context. Section~4 discusses the possible origin of the GeV emission and gives our conclusions.

\section{Data and Analysis} \label{sec:data}

\subsection{Fermi-LAT Analysis and Results} 
The \textit{Fermi} Large Area Telescope (LAT) provides continuous wide-field monitoring of the GeV $\gamma$-ray sky, making it well suited for long-term studies of source detection and variability \citep{Atwood2009LAT}. We analyzed the LAT all-sky survey data from 2008 August 4 (MJD 54682) to 2025 July 31 (MJD 60887) using \texttt{Fermitools} (v2.5.1). Events were selected with \texttt{evtype=3} and \texttt{evclass=128} in the energy range from 100 MeV to 500 GeV. Standard data quality cuts, \texttt{DATA\_QUAL>0 \&\& LAT\_CONFIG==1}, were applied with the \texttt{gtmktime} tool. We adopted a circular region of interest centered on 1H 1934$-$063 with a radius of $15^\circ$. The background model was constructed from the FL16Y source list \citep{Ballet2026FL16Y}, together with the Galactic diffuse component \texttt{gll\_iem\_v07} and the isotropic template \texttt{iso\_P8R3\_SOURCE\_V3\_v1}. The target source was modeled with a power-law spectrum.

A full-period likelihood fit with the FL16Y model gives only a modest excess at the position of 1H 1934$-$063, with TS = 12.41 in the 0.1--500 GeV band and TS = 14.77 in the 1--500 GeV band. We therefore focus on the time-resolved analysis below. To examine the long-term $\gamma$-ray behavior of the target region, we extracted binned light curves in both the 0.1--500 GeV and 1--500 GeV bands, using time bins of 1 yr and 0.5 yr, as shown in Figure~1. In each time bin, we performed an independent likelihood fit and derived a 95\% upper limit when TS $<9$. For comparison, we also show the TS evolution of the nearby FL16Y source J1936.9--0552. The test statistic is defined as
\begin{equation}
{\rm TS}=2\left(\log \mathcal{L}_{1}-\log \mathcal{L}_{0}\right),
\end{equation}
where $\mathcal{L}_{1}$ and $\mathcal{L}_{0}$ are the likelihoods with and without the source, respectively; for a point source, the detection significance is approximately $\sqrt{\rm TS}$. The most prominent excess toward 1H 1934--063 appears around MJD 58788--59031. The 0.5 yr light curves show that the enhancement remains confined to this interval within the available photon statistics, while FL16Y J1936.9--0552 does not show a corresponding significant enhancement. We therefore performed a flare-period analysis of MJD 58788--59031, including the TS and residual maps shown in Figure~2.

For the flare-period analysis, we adopted a circular region of
interest with a radius of $10^\circ$. An initial fit over
0.1--500~GeV, with both the normalization and photon index of the
target left free, yielded a TS value of approximately 27. However, the
integrated photon flux was poorly constrained because of the strong
degeneracy between the spectral normalization and photon index.
Increasing the lower energy threshold from 0.1 to 1~GeV left the TS
value nearly unchanged while substantially reducing the relative flux
uncertainty. We therefore adopted the 1--500~GeV analysis as the
primary result for the flare interval.

The 1--500~GeV TS map exhibits a clear excess near
1H~1934--063, while the residual map does not show any obvious
structured residuals at the target position. The likelihood fit gives
a TS value of 27.12, a photon index of
$\Gamma=1.50\pm0.25$, and a photon flux of
$(4.94\pm2.33)\times10^{-10}\,
\mathrm{ph\,cm^{-2}\,s^{-1}}$. A localization analysis with
\texttt{gtfindsrc} yields a best-fit position of
$(294.396241^\circ,-6.225778^\circ)$ and a 95\% statistical error
radius of approximately $0.061^\circ$. The radio position of
1H~1934--063 lies within this error circle, whereas the nearby source
FL16Y~J1936.9--0552 is not significantly detected during the same
interval.

We further evaluated the flux uncertainties using profile-likelihood
scans. For the 1--500~GeV fit, the 68\% profile-likelihood interval
corresponds to a photon flux of
$4.94^{+2.88}_{-2.08}\times10^{-10}\,
\mathrm{ph\,cm^{-2}\,s^{-1}}$. As a supporting wider-band
measurement, we repeated the 0.1--500~GeV fit with the photon index
fixed at the value derived from the 1--500~GeV analysis. This fit
gives TS $=26.91$ and a photon flux of
$1.62^{+0.80}_{-0.62}\times10^{-9}\,
\mathrm{ph\,cm^{-2}\,s^{-1}}$, where the uncertainties correspond to
the 68\% profile-likelihood interval. The larger integrated flux and
its larger absolute uncertainty mainly result from extending the same
power-law model to lower energies and do not imply that the low-energy
events dominate the detection. When the photon index is allowed to
vary, the inclusion of the low-energy data also strengthens the
normalization--index degeneracy, producing the poorly constrained
free-index result described above.

To examine the energy distribution of the events associated with the
localized excess, we applied \texttt{gtsrcprob} to the final
likelihood model. The high-probability events are concentrated in the
10--50~GeV band, where five photons have
$P_{\rm target}\geq0.8$, including four with
$P_{\rm target}\geq0.9$. No photon with
$P_{\rm target}\geq0.5$ is found outside this energy range. The
highest-energy high-probability event has an energy of 21.7~GeV, with
$P_{\rm target}=0.984$ and
$P_{\rm nearby}=4.6\times10^{-4}$. This distribution is consistent
with the likelihood signal being driven mainly by events in the
10--50~GeV band, rather than by the larger number of lower-energy
events. The large flux uncertainty therefore reflects both the limited number of high-probability source events and the resulting normalization--index degeneracy. These probabilities are model-dependent and are used here only as a photon-level diagnostic, rather than as an independent
association criterion.

\begin{figure}
 \centering
        \includegraphics[width=\linewidth]{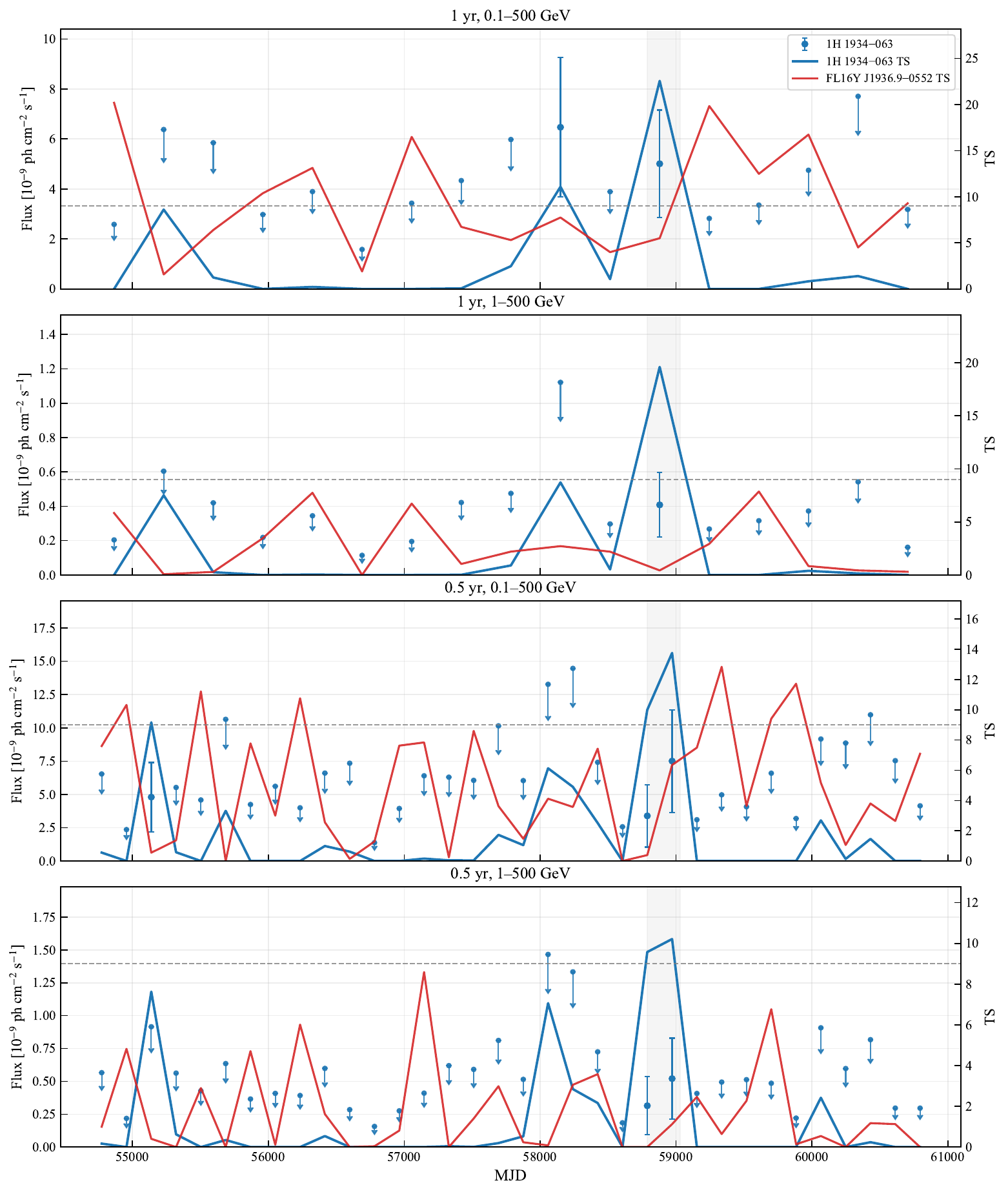}
 \caption{\textit{Fermi}-LAT light curves of 1H 1934--063 obtained with the FL16Y source model. 
The panels compare different energy ranges and temporal binnings, as labeled. 
Blue points and arrows denote the photon fluxes and 95\% upper limits of 1H 1934--063, while the blue and red curves show the TS values of 1H 1934--063 and FL16Y J1936.9--0552, respectively. 
Upper limits are shown for bins with TS $<9$. 
The dashed line marks TS = 9, and the gray shaded region indicates the flare interval MJD 58788--59031.}
\label{light_curve}
\end{figure}

\begin{figure*}
\centering
\includegraphics[width=\textwidth]{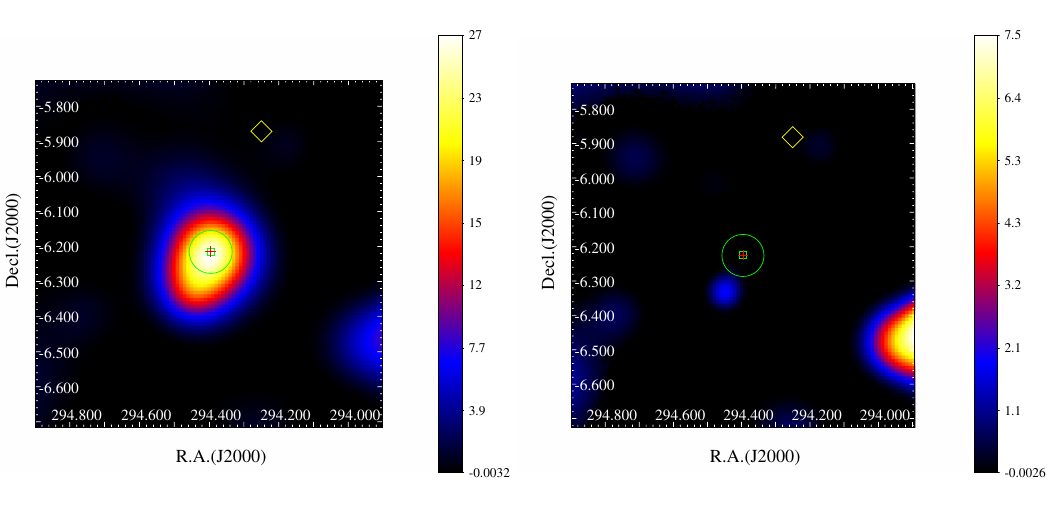}
 \caption{TS map (left) and residual map (right) in the 1--500 GeV band for the flare interval MJD 58788--59031. 
The green circle shows the 95\% statistical localization region centered on the best-fit $\gamma$-ray position of the excess. The red cross and green box indicate the optical and radio positions of 1H 1934--063, respectively, while the yellow diamond marks the nearby source FL16Y J1936.9--0552.}
\label{fig:tsmap}
\end{figure*}

\begin{figure}
 \centering
        \includegraphics[width=\linewidth]{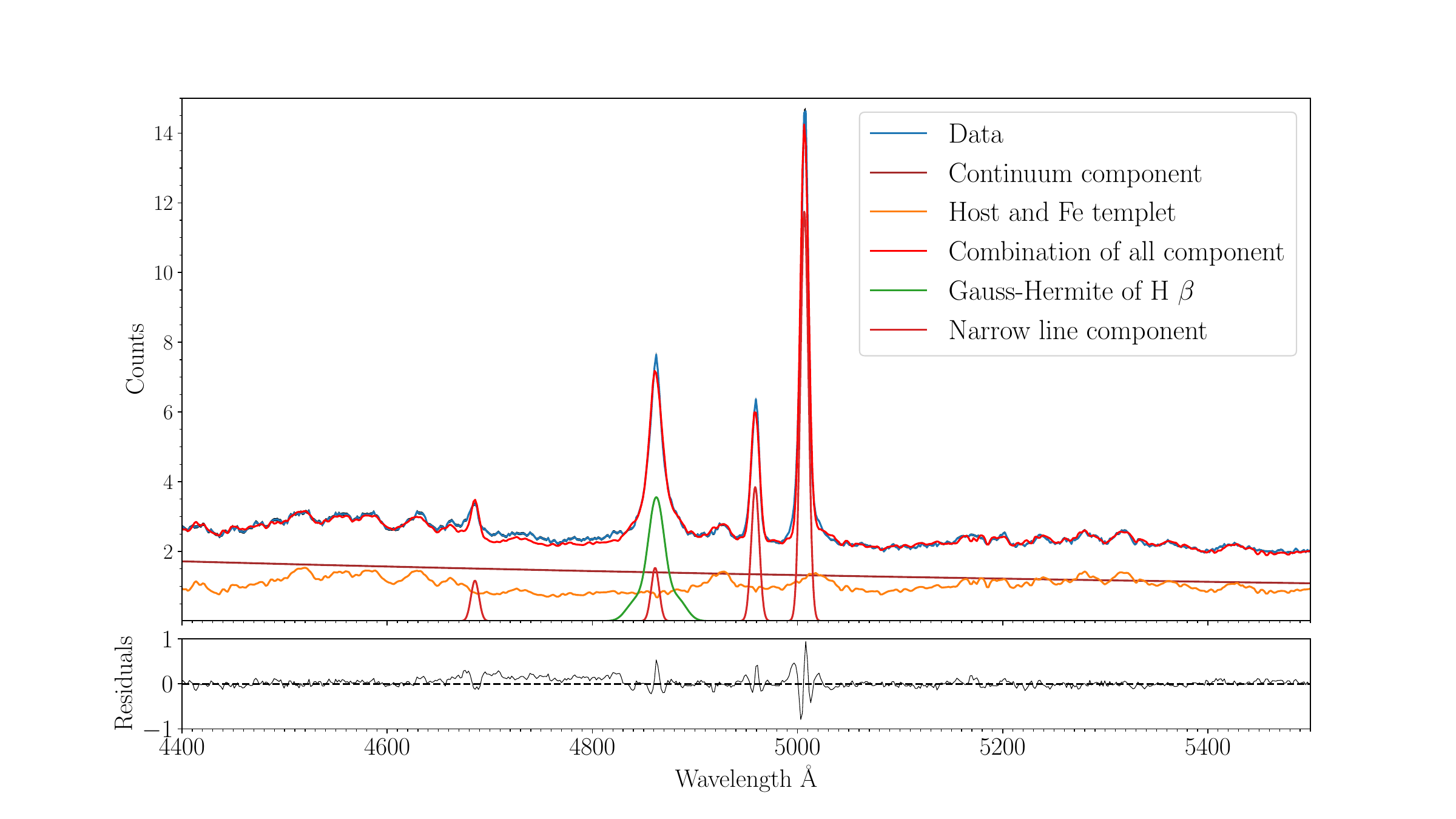}
 \caption{Spectral decomposition of the 6dFGS H$\beta$--[O\,III] region for 1H 1934$-$063. The upper panel shows the observed counts spectrum (blue), the continuum component, the host-galaxy and Fe template, the narrow-line component, the Gauss--Hermite model for the broad H$\beta$ component, and the total best-fit model. The lower panel shows the residuals of the fit. The decomposition indicates that the narrow-line contribution is non-negligible, while the broad H$\beta$ profile is better described by a Gauss--Hermite function than by a single Lorentzian.}
\label{fig:optical}
\end{figure}
\subsection{Optical Spectrum} 
Although 1H~1934$-$063 has previously been classified as an NLS1 \citep{Chen2018South}, we re-examined the 6dFGS spectrum in the H$\beta$--[O~III] region to verify its optical classification. NLS1s are commonly defined by ${\rm FWHM}({\rm H}\beta)<2000~{\rm km~s^{-1}}$ and $[{\rm O\,III}]\,\lambda5007/{\rm H}\beta<3$ \citep{Osterbrock1985,Goodrich1989}. Since the original 6dFGS spectrum is not flux calibrated, we focus on the local line profile and relative line strengths. The [O~III] emission indicates that the narrow H$\beta$ component cannot be neglected, so it is modeled explicitly. We find that the broad H$\beta$ profile is better described by a Gauss--Hermite function than by a single Lorentzian, yielding ${\rm FWHM}=1551\pm197~{\rm km~s^{-1}}$. The integrated line-flux ratio is $[{\rm O\,III}]\,\lambda5007/{\rm H}\beta=1.045$, well below the canonical NLS1 threshold. These measurements independently confirm the NLS1 classification of 1H~1934$-$063.
\section{Multiwavelength Properties of 1H~1934$-$063}

\begin{figure*}
\centering
\includegraphics[width=\textwidth]{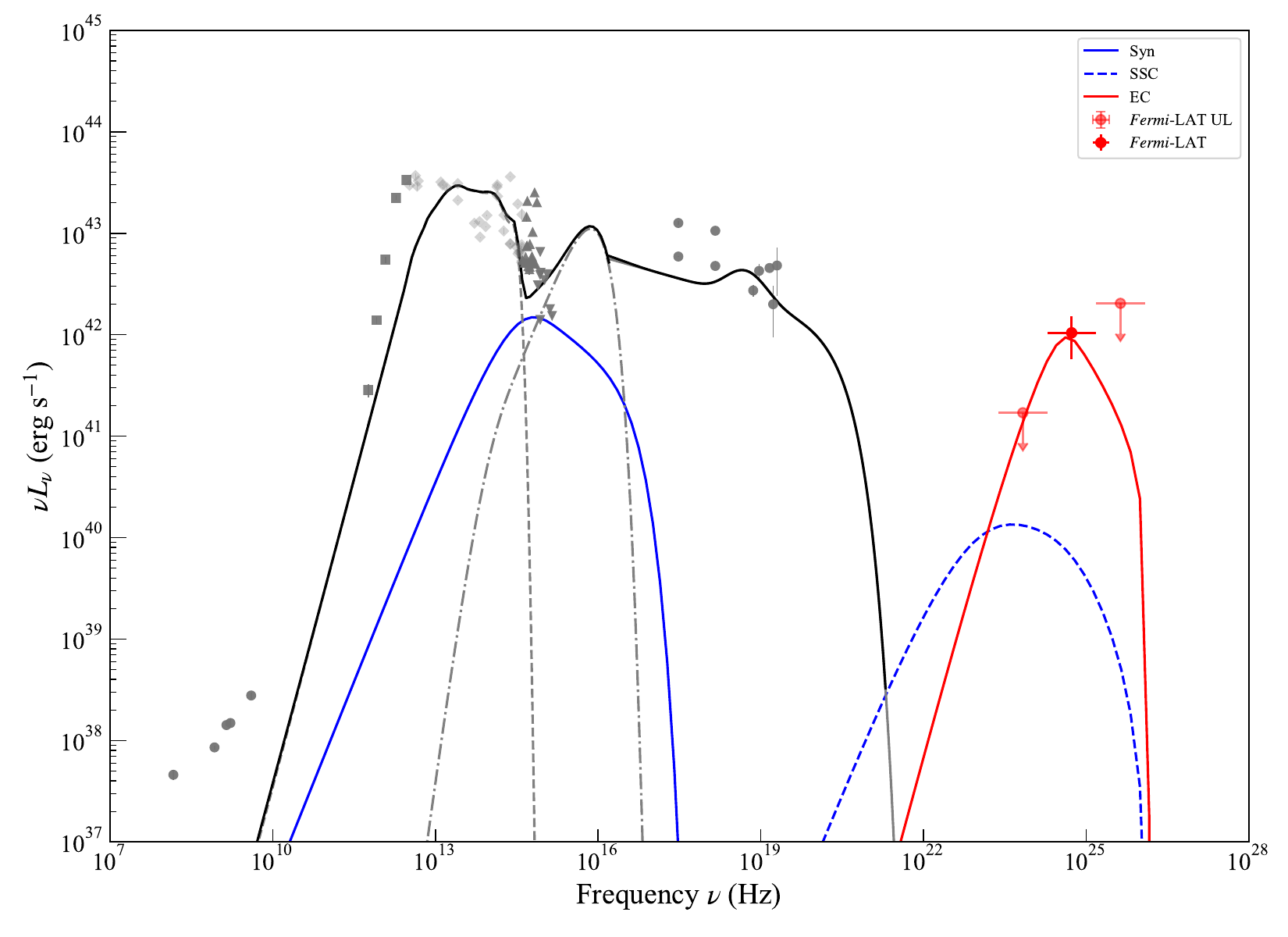}
\caption{Non-simultaneous broadband spectral energy distribution of 1H 1934$-$063. The gray points show archival radio-to-X-ray measurements compiled from the literature and public catalogs \citep{deGasperin2018SpecIndex,Franzen2021GLEAM,Mahony2010RASS6dFGS,Duchesne2025RACSHigh,Gordon2021VLASS,Ichikawa2019BASSXI,Melendez2014HerschelBAT,Abrahamyan2015IRASPSCFSC,Mauch2007SixdFGSRadio,Nagao2001OIII,Asmus2020LASr,Ishihara2010AKARI,Teplitz2010SpitzerSourceList,McMahon2013VHS,Hon2025BLAGN6dFGS,Cutri2003TwoMASS,Magnier2020PS1,Wolf2018SkyMapperDR1,Lasker2008GSC,Gaia2018DR2,Gaia2021EDR3,Gaia2023DR3,Henden2016APASS,Altmann2017HSOY,Stassun2018TIC,Page2012XMMSUSS,Chen2018South,Yershov2014UVSourceCatalogues,Boissay2016SoftExcess,Rodriguez2008IntegralSwift,Bird2007IBIS,Sazonov2007IntegralAGN,Malizia2009ComptonThick,Beckmann2006IntegralAGN,Cusumano2010BAT54m}; The red symbols denote the \textit{Fermi}-LAT detections and 95\% upper limits during the flare interval. 
The blue solid, blue dashed, and red solid curves show representative synchrotron, SSC, and EC components, respectively, 
while the black curve represents the phenomenological thermal contribution. 
The model curves are intended as an illustrative broadband representation rather than a unique physical fit.}
\label{fig:sed}
\end{figure*}

For the non-simultaneous broadband SED, we compiled archival measurements from public catalogs and the literature. The radio points are drawn from low- and mid-frequency survey products and radio identifications \citep{deGasperin2018SpecIndex,Franzen2021GLEAM,Mahony2010RASS6dFGS,Duchesne2025RACSHigh,Gordon2021VLASS}. The infrared through ultraviolet coverage combines all-sky or large-area source catalogs and source-specific compilations, including BASS, IRAS, AKARI, Spitzer source lists, 2MASS, VHS, Pan-STARRS1, SkyMapper, Gaia, APASS, XMM-SUSS, and the 6dFGS southern NLS1 sample \citep{Ichikawa2019BASSXI,Melendez2014HerschelBAT,Abrahamyan2015IRASPSCFSC,Mauch2007SixdFGSRadio,Nagao2001OIII,Asmus2020LASr,Ishihara2010AKARI,Teplitz2010SpitzerSourceList,McMahon2013VHS,Hon2025BLAGN6dFGS,Cutri2003TwoMASS,Magnier2020PS1,Wolf2018SkyMapperDR1,Lasker2008GSC,Gaia2018DR2,Gaia2021EDR3,Gaia2023DR3,Henden2016APASS,Altmann2017HSOY,Stassun2018TIC,Page2012XMMSUSS,Chen2018South,Yershov2014UVSourceCatalogues}. The X-ray measurements are taken from previous survey and source studies \citep{Boissay2016SoftExcess,Mahony2010RASS6dFGS,Rodriguez2008IntegralSwift,Bird2007IBIS,Sazonov2007IntegralAGN,Malizia2009ComptonThick,Beckmann2006IntegralAGN,Cusumano2010BAT54m}.

Because the broadband data are non-simultaneous, we model the infrared, optical--UV, and X-ray components only at a phenomenological level. The infrared continuum is represented by the infrared portion of the \citet{Elvis1994} quasar SED template. The optical--UV thermal emission is described with a multitemperature thin disk around a black hole of mass $M_{\rm BH}=8.1\times10^{6}\,M_\odot$, extending from $3$ to $3000$ Schwarzschild radii ($R_{\rm S}=2GM_{\rm BH}/c^{2}$), with a mass accretion rate of order $\dot{M}\sim10^{24}\ {\rm g\ s^{-1}}$ \citep{ShakuraSunyaev1973}. For the X-ray band, we adopt an empirical corona-plus-reflection form without an additional soft-excess component, since the X-ray spectrum of 1H 1934$-$063 is complex and highly variable and the data used here are not simultaneous \citep{Frederick2018,Xu2022}. These components are included only to sketch the thermal part of the broadband SED rather than to provide a detailed physical fit.

In the non-simultaneous broadband SED, the flare-period LAT
measurements show a significant detection in the
7.94--63.0~GeV band, with TS $\simeq 33.5$, while the
1.0--7.94~GeV and 63.0--500~GeV bands are represented by
95\% upper limits. Given the limited energy-bin coverage and the lack
of contemporaneous multiwavelength observations, the LAT measurements
do not uniquely determine the shape of the high-energy component. To
describe the nonthermal emission phenomenologically, we adopt a
one-zone leptonic scenario in which a broken-power-law electron
population radiates in a spherical emitting region\citep{Dermer1993,Ghisellini2009,Boettcher2013},

\begin{equation}
N(\gamma)=
\begin{cases}
K_{\rm e}\,\gamma^{-p_{1}}, & \gamma_{\min} \le \gamma < \gamma_{\rm b},\\[4pt]
K_{\rm e}\,\gamma_{\rm b}^{p_{2}-p_{1}}\,\gamma^{-p_{2}}, & \gamma_{\rm b} \le \gamma \le \gamma_{\max},
\end{cases}
\end{equation}
where $K_{\rm e}$ is the normalization and $p_{1}$ and $p_{2}$ are the low- and high-energy indices, respectively. A representative parameter set for the phenomenological synchrotron+SSC+EC model is listed in Table~\ref{tab:sed_params}. Within this framework, the synchrotron component describes the low-frequency nonthermal emission. For a radio-quiet source such as 1H 1934--063, this component is constrained by the modest observed radio luminosity and must remain subdominant in the radio band. The SSC contribution is also too weak and too broad to reproduce the LAT feature. We therefore use an additional EC component as a phenomenological representation of the high-energy emission, approximating the external radiation field by an equivalent blackbody with $T \sim 2\times10^4$ K and $u_{\rm ext}\sim10^{-4}\ {\rm erg\ cm^{-3}}$. In the Thomson limit, the characteristic observed EC frequency can be written approximately as
\begin{equation}
\nu_{\rm EC} \approx \frac{4}{3}\,\gamma^{2}\,\nu_{\rm ext}\,\frac{\delta}{1+z},
\end{equation}
which shows that an EC component peaking in the GeV band requires
electrons with comparatively high Lorentz factors. In the
representative model adopted here, a relatively narrow electron
distribution, with $\gamma_{\min}$ close to $\gamma_{\rm b}$, is used
to reproduce the hard GeV component while keeping the associated
synchrotron and SSC emission subdominant. Given the limited LAT
spectral constraints, the non-simultaneous broadband data, and the
radio-quiet nature of the source, this parameterization should be
regarded as illustrative rather than unique. The nonthermal component
is therefore used only as a phenomenological description of a compact
high-energy emitter and does not identify a specific emission channel.

\begin{table}
\caption{Parameters of the one-zone leptonic fit to the non-simultaneous broadband SED.}
\centering
\begin{tabular}{lc}
\hline
Parameter & Value \\
\hline
$K_{\rm e}$ & $60$ \\
$p_{1}$ & $2.0$ \\
$p_{2}$ & $3.8$ \\
$\gamma_{\min}$ & $9.3\times10^{3}$ \\
$\gamma_{\rm b}$ & $1.0\times10^{4}$ \\
$\gamma_{\max}$ & $1.0\times10^{5}$ \\
$B$ (G) & $0.05$ \\
$R$ (cm) & $3.0\times10^{16}$ \\
$\delta$ & $13$ \\
$T_{\rm ext}$ (K) & $2.0\times10^{4}$ \\
$u_{\rm ext}$ (erg cm$^{-3}$) & $1.0\times10^{-4}$ \\
\hline
\end{tabular}
\label{tab:sed_params}
\end{table}

\section{Discussion and Conclusions}

Our main observational result is a flare-period \textit{Fermi}-LAT excess spatially consistent with 1H~1934--063. The refined localization and the contemporaneous non-detection of FL16Y~J1936.9--0552 make
1H~1934--063 a plausible counterpart, although they do not establish a formal association. Both the model-dependent source-probability analysis and the binned SED indicate that the signal is concentrated in the $\sim8$--63~GeV range, with only upper limits in the adjacent energy bands. Given the sparse LAT spectral coverage and the non-simultaneous broadband data, the shape and origin of the high-energy component remain poorly constrained. In the representative one-zone leptonic model adopted here\citep{Dermer1993,Ghisellini2009,Boettcher2013}, the SSC component alone is insufficient, whereas the addition of an EC component provides an illustrative phenomenological description. This model should not be
interpreted as a unique physical solution.

Alternative origins should also be considered. A star-formation-dominated interpretation appears less natural as the primary origin of the LAT excess, because the enhancement is confined to a limited time interval, whereas the $\gamma$-ray emission of star-forming galaxies is generally expected to be comparatively steady
and to follow empirical scaling relations with infrared luminosity\citep{Ackermann2012SFG}. Nevertheless, in the absence of a source-specific infrared or star-formation-rate estimate, a steady contribution from the host galaxy cannot be quantitatively excluded. Recent LAT studies have also begun to reveal high-energy emission from radio-quiet Seyferts. A stacked analysis of nearby, ultrahard-X-ray-bright radio-quiet AGNs reported an average GeV signal that was interpreted in terms of coronal emission. A potential detection in gamma rays of the radio-quiet Seyfert NGC~3281 has been reported \citep{Liu2025NGC3281,Liu2025HotCoronae}, with a softer $\gamma$-ray component that cannot be explained by star formation alone and may arise from either the corona or a jet-related component. 

For 1H~1934--063, a coronal origin is physically motivated by its strong X-ray variability and reverberation signatures, which indicate an active, compact inner X-ray-emitting region \citep{Frederick2018}. However, these observations were not contemporaneous with the LAT-active interval and cannot establish a direct X-ray/$\gamma$-ray connection. Although nonthermal particles in AGN coronae can produce high-energy radiation, the intense disk and coronal photon fields are expected to make a canonical compact corona opaque to $\gamma$ rays through internal $\gamma\gamma$ absorption \citep{Inoue2019,Inoue2020}. The high-probability events at
10--50~GeV are therefore difficult to accommodate in such a region, although a coronal contribution cannot be excluded without a source-specific opacity calculation. An extended nonthermal coronal region at larger radii could reduce the target-photon density and permit the escape of photons above several GeV \citep{Liu2025HotCoronae}. This interpretation, however, remains population-based and model-dependent, and has not been established for 1H~1934--063. The coronal scenario therefore remains viable but unconfirmed, motivating consideration of a less compact emission region.

A weak or transient jet-related component offers one possible realization of such a less compact emission region. The radio properties of 1H~1934--063 do not, however, support a persistent, powerful, blazar-like jet. The source is radio quiet, with $R_{1.4}\simeq2.8$ and $R_{5.5}\simeq1.6$, and its catalog-level 1.4--5.5~GHz spectrum is steep. At 5.5~GHz, the emission is diffuse around the central component and broadly follows the host-galaxy morphology, rather than displaying a clearly resolved collimated jet \citep{Chen2018South, Chen2020Radio}. These measurements constrain the long-term radio output, but they are non-simultaneous and do not resolve the parsec-scale nuclear region. Moreover, compact nonthermal cores and weak parsec-scale jets have been observed in other radio-quiet Seyferts and NLS1s \citep{Doi2013,Panessa2019,Yao2021}. Radio quietness therefore constrains the strength and duty cycle of any jet rather than proving its absence. A compact, intermittent, or low-duty-cycle jet-related component remains possible, although confirming such a scenario would require contemporaneous high-frequency radio observations, VLBI imaging, variability measurements, or polarization constraints.

The ionized outflow reported in 1H~1934--063 highlights the complexity of its nuclear environment \citep{Xu2022}. More broadly, accretion-modified stars, embedded compact objects, and localized outflows or jets within AGN disks have been proposed as transient high-energy engines\citep{Wang2021AMSBBH,Liu2024AMSObs,Chen2023COAGN,Chen2025JetAGN}. However, no independent evidence links the flare-period $\gamma$-ray signal to a disk-embedded event, so these scenarios are considered only as broader theoretical context.

The principal limitation of this work is the lack of contemporaneous multiwavelength data. The archival infrared, optical--UV, and X-ray measurements used in the SED do not uniquely constrain the model parameters. Simultaneous multiwavelength observations, improved LAT photon statistics during future high states, and high-resolution radio imaging are needed to test coronal, weak-jet, and other localized high-energy scenarios. 1H~1934--063 therefore provides an interesting case for probing high-energy activity in radio-quiet NLS1 galaxies beyond the conventional radio-loud blazar framework.

\begin{acknowledgments}
We thank Dingrong Xiong, Shasha Li, and Haicheng Feng for constructive comments.
This research has made use of public data and analysis resources provided by the Fermi Science Support Center, as well as the NASA/IPAC Extragalactic Database (NED), available via \dataset[https://doi.org/10.26132/NED1]{https://doi.org/10.26132/NED1}. We acknowledge the National Key R\&D Program of China under Grant No.~2021YFA1600404.
\end{acknowledgments}

\facilities{Fermi(LAT)}

\software{Fermitools \citep{Fermitools_ascl} (v2.5.1)}

\bibliographystyle{aasjournal}
\bibliography{cite_0}
\end{document}